\documentclass[showpacs,showkeys,amsmath,amssymb,preprint,nofootinbib]{revtex4-1}
\usepackage{graphicx}
\usepackage{epsfig}
\usepackage{bm}
\usepackage{float}
\usepackage[T1]{fontenc}
\usepackage[utf8]{inputenc}
\usepackage{graphicx}
\usepackage{bm}
\usepackage{amsmath}
\usepackage{xcolor}
\usepackage{siunitx}
\usepackage[pdftex, pdftitle={Article}, pdfauthor={Author}]{hyperref}
\DeclareUnicodeCharacter{2212}{\textendash}
\def\beq{\begin{eqnarray}}

\def\eeq{\end{eqnarray}}

\usepackage{scalerel}
\usepackage{tikz}
\usetikzlibrary{svg.path}

\definecolor{orcidlogocol}{HTML}{A6CE39}
\tikzset{
  orcidlogo/.pic={
    \fill[orcidlogocol] svg{M256,128c0,70.7-57.3,128-128,128C57.3,256,0,198.7,0,128C0,57.3,57.3,0,128,0C198.7,0,256,57.3,256,128z};
    \fill[white] svg{M86.3,186.2H70.9V79.1h15.4v48.4V186.2z}
                 svg{M108.9,79.1h41.6c39.6,0,57,28.3,57,53.6c0,27.5-21.5,53.6-56.8,53.6h-41.8V79.1z M124.3,172.4h24.5c34.9,0,42.9-26.5,42.9-39.7c0-21.5-13.7-39.7-43.7-39.7h-23.7V172.4z}
                 svg{M88.7,56.8c0,5.5-4.5,10.1-10.1,10.1c-5.6,0-10.1-4.6-10.1-10.1c0-5.6,4.5-10.1,10.1-10.1C84.2,46.7,88.7,51.3,88.7,56.8z};
  }
}

\newcommand\orcidicon[1]{\href{https://orcid.org/#1}{\mbox{\scalerel*{
\begin{tikzpicture}[yscale=-1,transform shape]
\pic{orcidlogo};
\end{tikzpicture}
}{|}}}}

\begin{document}

\title[DEA]{Spatial curvature in Unimodular Gravity}

\author{Gilberto Aguilar-Pérez$^1$\orcidicon{0000-0001-6821-4564}}
\email{gilaguilar@uv.mx}

\author{Miguel Cruz$^1$\orcidicon{0000-0003-3826-1321}}
\email{miguelcruz02@uv.mx}

\author{Samuel Lepe$^2$\orcidicon{0000-0002-3464-8337}}
\email{samuel.lepe@pucv.cl}

\affiliation{$^1$Facultad de F\'{\i}sica, Universidad Veracruzana 91097, Xalapa, Veracruz, M\'exico,\\
$^2$Instituto de F\'\i sica, Pontificia Universidad Cat\'olica de Valpara\'\i so, Casilla 4950, Valpara\'\i so, Chile.}

\date{\today}

\begin{abstract}
We investigate the cosmological implications of unimodular gravity (UG) featuring energy diffusion and spatial curvature. While standard diffusion models often suffer from thermodynamic inconsistencies, we propose a phenomenologically viable power-law Ansatz for the diffusion function, $Q(z) = Q_0(1+z)^\beta$, which strictly satisfies the second law of thermodynamics by demanding positive entropy production ($\beta Q_0 > 0$). Using a joint statistical analysis with the Pantheon+ Type Ia Supernova compilation and Baryon Acoustic Oscillation (BAO) measurements, we tightly constrain the parameter space. We find a diffusion exponent of $\beta = 0.503_{-0.126}^{+0.118}$ and a slight preference for a closed spatial geometry with $\Omega_{k0} = -0.109_{-0.071}^{+0.076}$ at present time. Remarkably, the consideration of spatial curvature and diffusion naturally alleviates the Hubble tension, yielding $H_0 = 73.350_{-0.226}^{+0.221}$ km/s/Mpc while maintaining a consistent cosmic age of $t_0 \simeq 13.61$ Gyr. Furthermore, the constrained diffusion scales as a stable, quintessence-like effective dark energy ($\omega_{\text{eff}} \simeq -0.832$). Thus, unimodular diffusion provides a thermodynamically consistent phenomenological alternative that can alleviate the Hubble tension while preserving both the cosmic age and the sound-horizon scale, with a preference for a closed spatial geometry.
\end{abstract}



\maketitle

\section{Introduction}
\label{sec:intro}

At the core of the standard cosmological paradigm lies the assumption of strict local conservation of the energy-momentum tensor, a direct consequence of the full diffeomorphism invariance of General Relativity. While this foundational framework has provided a highly successful macroscopic description of the Universe, it links vacuum energy to spacetime curvature, giving rise to the unresolved cosmological constant problem \cite{weinberg}. A profound and theoretically elegant alternative is UG. By restricting the symmetries of the gravitational action to volume-preserving diffeomorphisms, UG naturally decouples the vacuum energy from the geometric background, allowing the cosmological constant to emerge simply as an integration constant.

Beyond its fundamental theoretical appeal, this restricted symmetry introduces a transformative phenomenological consequence: the energy-momentum tensor is no longer required to be strictly conserved. Instead, its divergence acts as a source for an effective energy diffusion process, $Q(t)$, which can dynamically emulate the effects of dark energy and drive late-time cosmic acceleration, a feature that has been robustly tested against background cosmological constraints \cite{norman}. In this phenomenological description, the diffusion function should be interpreted as a macroscopic manifestation of energy exchange induced by the non-conservation of the energy-momentum tensor, rather than as a microscopic particle diffusion mechanism. Crucially, unlike the standard $\Lambda$CDM paradigm, which assumes a strictly adiabatic expansion ($dS=0$), the time-dependent diffusion in UG naturally induces non-adiabatic dynamics. This transition from a reversible to an irreversible process provides a more realistic physical description of cosmic evolution, as departures from equilibrium and dissipative effects are generally expected in complex systems. By treating expansion as a non-adiabatic process, UG can emulate complex mechanisms such as bulk viscosity or matter creation \cite{prigogine} without the need for additional fields or higher-curvature corrections, significantly simplifying the theory while providing a mechanism to address modern observational anomalies.

The advent of high-precision cosmology has dramatically highlighted the need to investigate such dynamical extensions. The $\Lambda$CDM model is currently challenged by severe anomalies, most notably the $>5\sigma$ Hubble ($H_0$) tension between early- and late-universe probes \cite{agha, riess}, as well as recent hints of a closed spatial geometry inferred from the Planck 2018 cosmic microwave background (CMB) data \cite{planck}. While energy diffusion provides a rich phenomenology to address these crises—as previously demonstrated by studies utilizing diffusion models in UG to alleviate the $H_{0}$ tension \cite{ugfran, ug}-its physical viability is fundamentally contingent upon thermodynamic consistency. Previous investigations into UG have often employed diffusion functions—such as the barotropic model ($Q \propto \rho$)—that lead to thermodynamic pathologies, including negative entropy production or violations of the Nernst theorem. To construct a physically rigorous model, it is imperative to enforce the second law of thermodynamics from the outset, demanding an intrinsically positive entropy production for the cosmic fluid.

In this work, we present a thermodynamically consistent cosmological model within the UG framework, characterized by a redshift-dependent power-law diffusion function and the inclusion of spatial curvature. We explicitly show that imposing thermodynamic stability via the condition $\beta Q_0 > 0$ naturally disfavors phantom-induced future singularities. By incorporating a non-zero curvature parameter $\Omega_{k0}\equiv\Omega_k(z=0)$, we prevent the artificial biasing of diffusion effects and provide a mechanism to alter geometric distance projections. 

Using the latest Pantheon+ Supernova dataset and comprehensive BAO measurements, we place stringent MCMC constraints on this model. As we will show, the interplay between energy non-conservation and spatial curvature significantly alleviates the $H_0$ tension while successfully preserving the age of the Universe and the integrity of the sound horizon scale.

The paper is organized as follows: In Section \ref{sec:unimodular}, we review the foundations of UG and the thermodynamic diffusion Ansatz. In Section \ref{sec:v}, we present the MCMC analysis and discuss the alleviation of the $H_0$ tension alongside cosmic age consistency. Finally, in Section \ref{sec:vi}, we provide our concluding remarks. In Appendix \ref{sec:gpapp}, we provide an independent consistency check of the diffusion function using a Gaussian Process reconstruction. In this study, we adopt units where $8\pi G = c = k_{B} = 1$, and our analysis is carried out within the framework of a Friedmann-Lema\^{i}tre-Robertson-Walker (FLRW) universe.
 
\section{Unimodular Gravity Framework}
\label{sec:unimodular}

In the framework of UG, the gravitational field is described by the trace-free part of the Einstein field equations \cite{ellis1, ellis2}
\begin{equation}
    R_{\mu\nu} - \frac{1}{4}g_{\mu\nu}R = T_{\mu\nu} - \frac{1}{4}g_{\mu\nu}T,\label{eq:eom}
\end{equation}
where $T$ represents the trace of the energy-momentum tensor. To obtain the field equations (\ref{eq:eom}) a constraint is introduced into the Einstein-Hilbert action. This equation of motion can be derived from various diffeomorphism-invariant action principles \cite{montesinos}. Specifically, the unimodular condition—which fixes the volume element to a specific functional form, $\sqrt{-g} = \zeta(x)$—is enforced through the use of a Lagrange multiplier field, $\lambda(x)$. The corresponding gravitational Lagrangian density takes the following form
\begin{equation}
    \mathcal{L} = \sqrt{-g} \left[ R + \lambda(x) \left( 1 - \frac{\zeta(x)}{\sqrt{-g}} \right) \right],
\end{equation}
where $R$ is the Ricci scalar and $\zeta(x)$ is a scalar density. This formulation is consistent with recent developments in diffeomorphism-invariant action principles for trace-free gravity, which ensure the recovery of the unimodular constraint through the variation of the action \cite{montesinos}. Under this formulation, the variation of the action with respect to the Lagrange multiplier $\lambda(x)$ directly recovers the unimodular condition $\sqrt{-g} = \zeta(x)$. Conversely, performing the variation with respect to the inverse metric $g^{\mu\nu}$ leads to the modified field equations
\begin{equation}
    R_{\mu\nu} - \frac{1}{2}g_{\mu\nu}R + \lambda(x)g_{\mu\nu} = T_{\mu\nu}.\label{eq:eom2}
\end{equation}
It is important to note that, unlike standard General Relativity where the cosmological constant is a fixed parameter in the Lagrangian, here $\lambda(x)$ emerges as a dynamical field that, upon using the Bianchi identities, is determined to be an integration constant plus a diffusion term, as we shall see. This shift in the role of $\lambda(x)$ from a fundamental constant to a Lagrange multiplier is what allows UG to potentially address the vacuum energy problem by decoupling the energy density of the vacuum from spacetime. By comparing Eqs. (\ref{eq:eom}) and (\ref{eq:eom2}), we can deduce that $\lambda(x)=(1/4)(R+T)$. A key implication of this framework, obtained by applying the Bianchi identities to (\ref{eq:eom}), is the possible violation of energy-momentum conservation
\begin{equation}
    \nabla^{\mu}T_{\mu\nu} = \frac{1}{4}\partial_{\nu}(R+T) \equiv J_{\nu},
\end{equation}
being $J_{\nu}$ the current for the non-conservation of the energy-momentum tensor. As demonstrated in foundational studies, such violations of energy conservation—even if originating at fundamental scales—can naturally source an effective dark energy component macroscopically \cite{josset}. This relation allows for the emergence of an integration constant $\Lambda$ and a diffusion function $Q(x)$, such that the Lagrange multiplier associated with the unimodular constraint can be identified as an effective cosmological constant
\begin{equation}
    \lambda_{\text{eff}}(x) = \Lambda + \int J(x)= \Lambda + Q(x).\label{eq:lag}
\end{equation}
being $\Lambda$ an arbitrary integration constant and $Q(x)$ the energy diffusion function emerging from the non-conservation of the energy-momentum tensor. This formula follows directly from the geometric properties of the trace-free Einstein tensor together with the Bianchi identity. Crucially, this relationship remains invariant under the inclusion of spatial curvature, as it does not depend on the specific value of $k$ but on the conservation properties of the geometric background. The inclusion of the diffusion term $Q(x)$ distinguishes UG from standard General Relativity with a cosmological constant, providing a dynamical mechanism to account for violations of energy conservation.

\subsection{Spatial Curvature}
\label{sec:curvature_motivation}

While the standard $\Lambda$CDM model typically assumes a spatially flat universe ($k=0$) motivated by the simplest inflationary paradigms, the assumption of spatial flatness has come under renewed scrutiny in light of recent high-precision observations. For instance, the latest photometric and spectroscopic measurements of high-redshift galaxies from the JWST have opened new frontiers to jointly test dark energy models and spatial curvature at early cosmic epochs \cite{jwst}. Notably, analyses of the Planck 2018 CMB power spectra have hinted at a preference for a closed geometry ($\Omega_k < 0$), a result that stands in tension with certain BAO datasets \cite{planck}. These indications suggest that spatial curvature may play a non-trivial dynamical role in late-time cosmology, acting as a catalyst for cosmic expansion \cite{ksing}. This {\it curvature tension} highlights the importance of treating $\Omega_k$ as a free parameter to avoid potential biases in the estimation of other cosmological observables; as demonstrated by Clarkson et al.; incorrectly assuming a flat geometry can induce critically large, redshift-dependent errors when reconstructing the dynamics of the dark sector \cite{clarkson}. In the framework of UG, the inclusion of spatial curvature is particularly relevant due to its coupling with the energy diffusion function. Since the effective cosmological constant $\lambda_{\text{eff}}(x)$ and the scaling of the matter density are modified by the diffusion term, spatial curvature provides an additional degree of freedom that is essential for breaking degeneracies within the modified expansion law, as it is well established that observational constraints often exhibit a strong degeneracy between spatial curvature and dark energy dynamics \cite{ryan}. The necessity of this extra degree of freedom has been recently highlighted by the use of JWST high-redshift galaxy data to simultaneously constrain spatial geometry and dark energy dynamics \cite{jwst}. Since the diffusion term already modifies the expansion rate, ignoring curvature may cause diffusion effects to be spuriously overestimated to compensate for geometric mismodeling. By introducing the spatial curvature parameter, we can more accurately distinguish between effects arising from the non-conservation of the energy-momentum tensor and those originating from the global geometry of spacetime. Additionally, the impact of spatial curvature on cosmological observables is primarily geometric, as it alters the relationship between redshift and distance. For instance, the transverse comoving distance, $D_{M}(z)$, which is fundamental for interpreting both BAO and Type Ia Supernova (SN Ia) data, is highly sensitive to the curvature parameter
\begin{equation}
D_M(z)=
\begin{cases}
\chi(z), & \Omega_{k0}=0, \\[0.2cm]
\dfrac{c/H_0}{\sqrt{\Omega_{k0}}}\sinh\left[\sqrt{\Omega_{k0}}\dfrac{H_0 \chi(z)}{c}\right], & \Omega_{k0}>0, \\[0.35cm]
\dfrac{c/H_0}{\sqrt{|\Omega_{k0}|}}\sin\left[\sqrt{|\Omega_{k0}|}\dfrac{H_0 \chi(z)}{c}\right], & \Omega_{k0}<0.
\end{cases}
\end{equation}
where $\chi(z)$ is the comoving distance. Consequently, a non-zero curvature shifts the predicted position of the acoustic peaks in the BAO spectrum and modifies the luminosity distance $d_{L}(z) = (1+z)D_{M}(z)$, thereby affecting the theoretical distance modulus $\mu(z)$ used to fit supernova light curves. As usual, we denote the cosmological redshift by $z$, with its connection to the scale factor (and to cosmic time) expressed through the standard relation $1+z = a^{-1}(t)$ with the present value $a_0=1$. To maintain homogeneity and isotropy of spacetime, the energy diffusion function is taken to be a function solely of cosmic time, $Q(x) \mapsto Q(t)$.

\subsubsection{Modified Friedmann Equations with Spatial Curvature}
\label{sec:friedmann_equations}

In the context of a FLRW universe with spatial curvature $k$ and a single fluid description, the field equations of UG lead to a modified expansion history. Considering a late-time scenario in which the integration constant is set to $\Lambda=0$, the first Friedmann equation emerging from (\ref{eq:eom2}) and (\ref{eq:lag}) is expressed as
\begin{equation}
    3H^{2}(t) = \rho(t) + Q(t) - \frac{3k}{a^2(t)},\label{eq:fried1}
\end{equation}
where $H(t)\equiv \dot{a}/a$ is the Hubble rate and $\rho(t)$ is the energy density of the cosmic fluid, since we consider a perfect fluid description for the energy-momentum tensor. In our notation, the dots stand for derivatives with respect to cosmic time. The term containing $k$ accounts for the geometric effect of spatial curvature and varies inversely with the square of the scale factor. To fully characterize the cosmic evolution in this framework, we must supplement the Friedmann equation with the acceleration and continuity equations. The spatial sector of the equations of motion given by (\ref{eq:eom2}) and (\ref{eq:lag}) takes the form
\begin{equation}
    2\dot{H} + 3H^2 = - \left(p(t)-Q(t) \right)+\frac{k}{a^2(t)},\label{eq:fried2}
\end{equation}
being $p(t)$ the cosmic fluid pressure. The interplay between the energy density, the diffusion function, and the geometric curvature term defines the complete background history used in our observational constraints. The non-conservation of the energy-momentum tensor is directly reflected in the continuity equation, the evolution of the energy density is governed by
\begin{equation}
    \dot{\rho} + 3H(\rho+p) = -\dot{Q},\label{eq:non}
\end{equation}
where we have used the Friedmann and acceleration equations (\ref{eq:fried1}) and (\ref{eq:fried2}). This expression highlights that the diffusion term $Q(t)$ acts as a source or sink for the matter sector.

\subsection{Thermodynamic Consistency and the Diffusion Ansatz}
\label{sec:thermo}
The physical viability of a diffusion model in UG is strictly governed by its thermodynamic behavior. In a FLRW universe, the non-adiabatic evolution of the cosmic fluid is characterized by the Gibbs relation  $TdS = dU+pdV$, where $V=V_{0}a^3{(t)}$ is the volume, $T$ is the temperature of the fluid and $U = \rho V$ its internal energy. Evaluating the time derivative in the Gibbs relation and considering (\ref{eq:non}), we can write the entropy production in terms of the redshift as follows \cite{thermo}
\begin{equation}
    T \frac{dS}{dz} = -V \frac{dQ}{dz}.
\end{equation}
To satisfy the second law of thermodynamics, the model must ensure positive entropy production $dS/dt > 0$, which translates to the condition $dS/dz < 0$ when expressed in terms of the redshift, implying $dQ/dz > 0$ since $V>0$ always. Previous investigations have demonstrated that standard diffusion models, such as the barotropic model ($Q = \alpha \rho$) or Continuous Spontaneous Localization (CSL) models, often lead to thermodynamic inconsistencies, including negative entropy production or violations of the Nernst theorem in certain parameter regions \cite{thermo, thphan}. To address these limitations, we adopt a power-law Ansatz proposed in \cite{thphan} for the diffusion function
\begin{equation}
    Q(z) = Q_0(1+z)^{\beta}, \label{eq:ansatz}
\end{equation}
where $Q_0$ represents its present-day value and $\beta$ is the scaling exponent. As is evident, under the assumptions of homogeneity and isotropy, the energy diffusion function is taken to depend only on cosmic time (or, equivalently, on redshift). This specific form ensures a consistent thermodynamic evolution, as the requirement for positive entropy production is reduced to the parameter constraint 
\begin{equation}
    \beta Q_0 > 0. \label{eq:cons}
\end{equation}
By enforcing this condition, the model functions solely as an energy source for the cosmic fluid, effectively avoiding the thermodynamic pathologies found in earlier diffusion frameworks while maintaining a well-defined physical region for the energy density and the Hubble parameter.

\section{Observational analysis}
\label{sec:v}
In this section, we constrain the diffusive dark matter model using a joint analysis of Baryon Acoustic Oscillation (BAO) measurements and the Pantheon+ Type Ia Supernova compilation.

\subsection{Model and background expansion}
We consider a simplified version of the unimodular-inspired diffusion model in which the cosmic fluid behaves as dust-like matter
($\omega = 0$) and the cosmological constant is set to zero, $\Lambda = 0$. This choice is motivated by the fact that, within UG the cosmological constant appears as an integration constant and may be treated independently of the matter-diffusion sector. The modified Friedmann equation (\ref{eq:fried1}) is written as 
\begin{equation}
3H^2(z)=\rho(z)+Q_0(1+z)^\beta-3k(1+z)^2,
\end{equation}
where $Q_0$ is the present amplitude of the diffusion term, $\beta$ controls its redshift scaling and $k$ is the spatial curvature constant. For $\omega=0$, together with Eqs. (\ref{eq:non}) and (\ref{eq:ansatz}) the matter density becomes
\begin{equation}
\rho(z)=\rho_0(1+z)^3\left\{1+\frac{\beta Q_0}{\rho_0(\beta-3)}
\left[1-(1+z)^{\beta-3}\right]\right\}.
\end{equation}
Introducing the dimensionless Hubble function
\begin{equation}
E(z)\equiv \frac{H(z)}{H_0},
\end{equation}
together with the density parameters at present time
\begin{equation}
\Omega_0=\frac{\rho_0}{3H_0^2},\qquad
\Omega_{Q0}=\frac{Q_0}{3H_0^2},\qquad
\Omega_{k0}=-\frac{k}{H_0^2},
\end{equation}
the expansion law can be written as
\begin{equation}
E^2(z)=
\left(\Omega_0+\frac{\beta\Omega_{Q0}}{\beta-3}\right)(1+z)^3
-\frac{3\Omega_{Q0}}{\beta-3}(1+z)^\beta
+\Omega_{k0}(1+z)^2.
\label{eq:Ez_diffusive}
\end{equation}
where we can identify three distinct contributions to the cosmic expansion: a term associated with the matter sector, a dark energy component arising from diffusion effects, and the spatial curvature, respectively. Since $\Lambda=0$, the normalization condition at $z=0$ is
\begin{equation}
1=\Omega_0+\Omega_{Q0}+\Omega_{k0},
\end{equation}
so that one of the density parameters is not independent. In our implementation, $\Omega_0$ is reconstructed through
\begin{equation}
\Omega_0 = 1-\Omega_{Q0}-\Omega_{k0}.
\end{equation}
Therefore, the free parameter set sampled in the MCMC analysis is
\begin{equation}
\Theta = \{H_0,\Omega_{Q0},\Omega_{k0},\beta\}.
\end{equation}
A singular behavior appears when $\beta=3$, which must be excluded from the allowed parameter space.

\subsection{Cosmological observables}
The background observables used in the likelihood are derived from the expansion history in Eq.~(\ref{eq:Ez_diffusive}).
The comoving distance is
\begin{equation}
\chi(z)=\frac{c}{H_0}\int_0^z \frac{dz'}{E(z')}.
\end{equation}
As mentioned above, the transverse comoving distance $D_M(z)$ depends on curvature:
\begin{equation}
D_M(z)=
\begin{cases}
\chi(z), & \Omega_{k0}=0, \\[0.2cm]
\dfrac{c/H_0}{\sqrt{\Omega_{k0}}}\sinh\left[\sqrt{\Omega_{k0}}\dfrac{H_0 \chi(z)}{c}\right], & \Omega_{k0}>0, \\[0.35cm]
\dfrac{c/H_0}{\sqrt{|\Omega_{k0}|}}\sin\left[\sqrt{|\Omega_{k0}|}\dfrac{H_0 \chi(z)}{c}\right], & \Omega_{k0}<0.
\end{cases}
\end{equation}
From these quantities, we construct the observables used in this work, namely, the Hubble distance:
\begin{equation}
D_H(z) = \frac{c}{H(z)}.
\end{equation} and the volume-averaged distance:
\begin{equation}
D_V(z) = \left[ z D_H(z) D_M^2(z) \right]^{1/3}.
\end{equation}
For supernovae, the luminosity distance is
\begin{equation}
d_L(z)=(1+z)D_M(z),
\end{equation}
and the theoretical distance modulus is
\begin{equation}
\mu(z)=5\log_{10}\left[\frac{d_L(z)}{\mathrm{Mpc}}\right]+25.
\end{equation}

\subsection{Datasets}

For Type Ia Supernovae we use the Pantheon+ compilation and its corresponding covariance matrix \cite{Brout2022,PantheonPlusData}.
For BAO we use a compilation including measurements from 6dFGS, SDSS Main Galaxy Sample, BOSS DR12, eBOSS, and DESI \cite{Beutler2011,Ross2015,Alam2017,Alam2021,DESI2024}. The BAO likelihood is constructed from the measured quantities
\begin{equation}
D_M(z)/r_d,\qquad D_H(z)/r_d,\qquad D_V(z)/r_d,
\end{equation}
where $r_d$ denotes the sound horizon at the drag epoch.

\subsection{Parameter constraints}

\begin{figure}[htbp]
\centering
\includegraphics[width=0.80\textwidth]{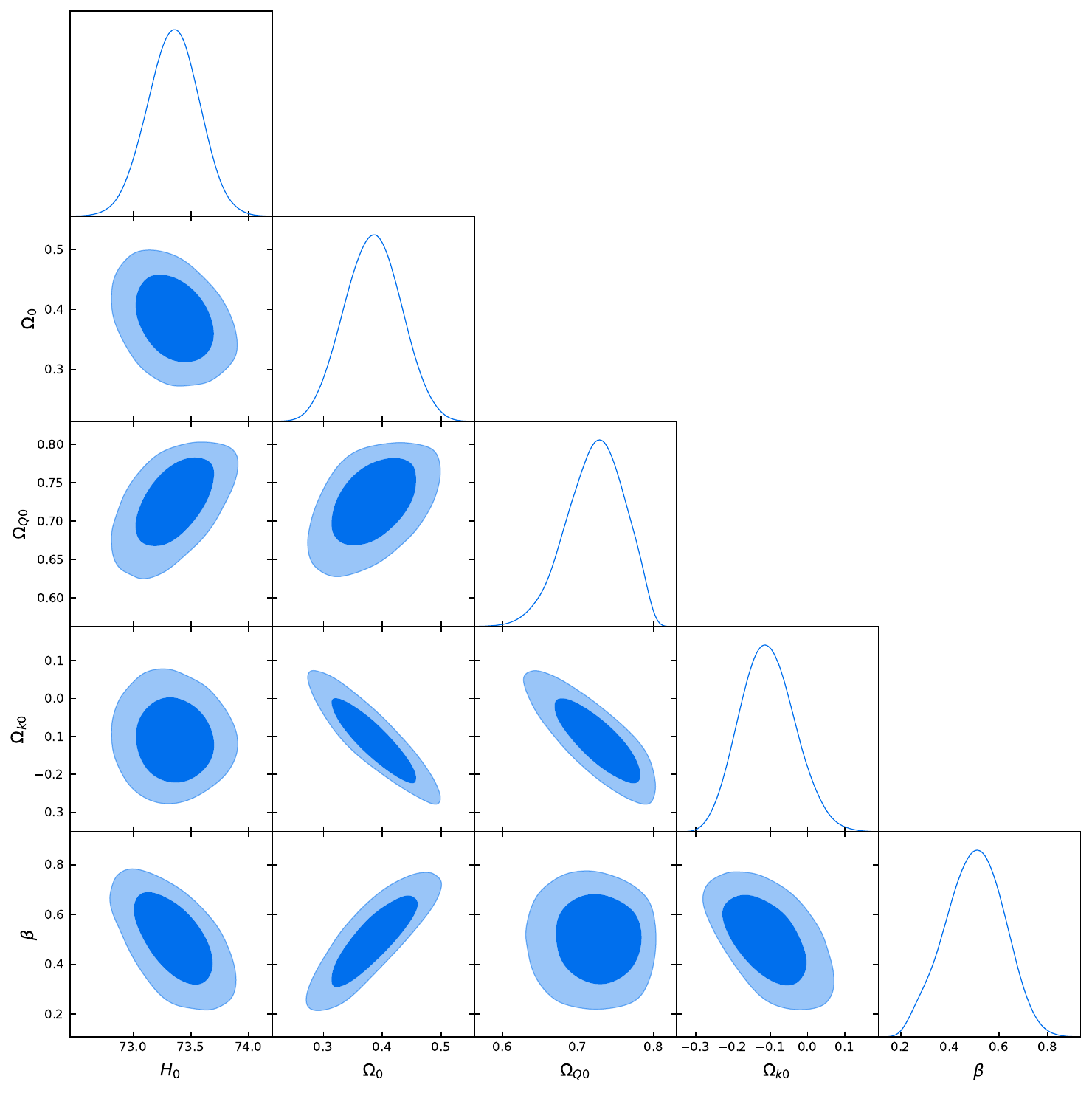}
\caption{Comparison of the 1D and 2D marginalized posterior distributions of the parameters $\beta$, $H_{0}$, $\Omega_0$, $\Omega_{Q0}$ and $\Omega_{k0}$.}
\label{fig:triangle}
\end{figure}

 
\begin{table}[h!]
\centering
\label{tab:bestfit_diffusive}
\begin{tabular}{lc}
\hline\hline
Parameter & Constraint \\
\hline
$H_0$ \mbox{(km/s/Mpc)} & $73.350^{+0.221}_{-0.226}$ \\
$\Omega_0$ & $0.384^{+0.048}_{-0.048}$ \\
$\Omega_{Q0}$ & $0.725^{+0.037}_{-0.040}$ \\
$\Omega_{k0}$ & $-0.109^{+0.076}_{-0.071}$ \\
$\beta$ & $0.503^{+0.118}_{-0.126}$ \\
\hline\hline
\end{tabular}
\caption{Marginalized constraints for the diffusive dark matter model.}
\end{table}
Using the cosmological observables discussed before, the obtained constraints for the diffusion exponent, $\beta = 0.503_{-0.126}^{+0.118}$, carry profound physical implications regarding the nature of the dark sector. By interpreting the energy diffusion function $Q(z)$ as an effective dark energy, we can describe the behavior of this component using an effective parameter state, $\omega_{\text{eff}} = \beta/3 - 1$. In other words, the exponent of the middle term on the right-hand side of Eq. (\ref{eq:Ez_diffusive}) is matched to the standard scaling form $3(1+\omega_{\text{eff}})$, which corresponds to dark energy with a constant parameter state $\omega_{\text{eff}}$. In this framework, the transition to a phantom regime ($\omega_{\text{eff}} < -1$) would require a negative scaling exponent, $\beta < 0$. However, as established by our thermodynamic consistency requirement in Eq. (\ref{eq:cons}), the second law of thermodynamics strictly demands that $\beta Q_{0} > 0$. Given that the MCMC analysis confirms a positive diffusion today, $\Omega_{Q0} = 0.725_{-0.040}^{+0.037} > 0$, the phantom regime remains outside the thermodynamically allowed region of the model. With an effective $\omega_{\text{eff}} \simeq -0.832$, the diffusion term acts as a stable, quintessence-type contribution, which naturally avoids future Big Rip-type singularities. This contrasts with other geometric dark energy frameworks, such as certain holographic cutoffs, where spatial curvature can actively accelerate the onset of a finite-time future singularity \cite{ksing}. Our result is in agreement with the most recent DESI findings, which indicate a preference for a time-varying dark energy component that currently exhibits quintessence-like behavior \cite{desi2, desi3}. Furthermore, the specific value of $\beta \approx 0.5$ allows us to break a potential geometric degeneracy that arises in modified expansion laws. It is noteworthy that if the diffusion exponent were $\beta = 2$, the diffusion term $Q(z)$ would scale exactly as $(1+z)^2$, mimicking the geometric contribution of the spatial curvature term $\Omega_{k0}(1+z)^2$ in the Friedmann equation. Such a case would imply that energy injection from the non-conservation of the energy-momentum tensor is observationally indistinguishable from intrinsic spatial curvature at the background level. The fact that our results constrain $\beta$ to a value significantly lower than $2$—and distinct from the standard matter scaling of $3$—demonstrates that the combined Pantheon+ and BAO data possess sufficient sensitivity to distinguish the dynamical effects of energy diffusion from the global geometry of spacetime.

The posterior distributions obtained from the MCMC analysis are shown in Fig.~\ref{fig:triangle}. The parameter constraints indicate a well-defined posterior region for the diffusion model. In particular, the diffusion exponent $\beta$ is constrained around values significantly below unity, while the curvature sector remains close to spatial flatness within uncertainties. The inferred values also show non-trivial correlations among $\Omega_0$, $\Omega_{Q0}$, and $\Omega_{k0}$, reflecting the normalization condition and the coupled structure of the modified expansion law. It is worth noting that the magnitude of the constrained curvature parameter, $|\Omega_{k0}| \simeq 0.109$, is notably higher than the stringent limits typically reported by the Planck collaboration when assuming a standard fiducial cosmology. This difference is a direct and expected consequence of our theoretical construction. In the standard paradigm, the dark energy sector is entirely rigid ($\omega=-1$), which forces the parameter estimation to heavily suppress any spatial curvature in order to fit the observational data. In contrast, by interpreting the dark energy sector as an effective, dynamically varying term that originates from energy diffusion, the rigid geometric bounds are inherently relaxed. The dynamical scaling of the diffusion function $Q(z)$ allows the parameter space to naturally accommodate a more pronounced closed geometry. Consequently, within the unimodular diffusion framework, this higher absolute value of curvature is not an anomaly, but rather the appropriate phenomenological region required to optimally balance the non-adiabatic expansion history with the geometric distance projections. On the other hand, our findings align with those reported in \cite{pavcur}, where the impact of the curvature parameter is analyzed in a thermodynamic framework. The main result shows that the generalized second law is always satisfied in a homogeneous and isotropic universe whose equation of state is not below -1, provided the spatial geometry is either flat or closed, whereas this is not guaranteed in universes with hyperbolic spatial sections.

To further assess the statistical performance of the proposed scenario, we compare the diffusive dark matter model against a curved $\Lambda$CDM cosmology using information criteria. The comparison is performed through the Akaike Information Criterion (AIC) and the Bayesian Information Criterion (BIC), which quantify the balance between goodness of fit and model complexity \cite{liddle}. For the diffusive model, we obtain
\begin{equation}
\Delta {\rm AIC} = -9.65,
\qquad
\Delta {\rm BIC} = -4.21,
\end{equation}
where the differences are defined relative to the curved $\Lambda$CDM reference model. According to standard interpretation scales, the AIC result indicates strong evidence in favor of the diffusive scenario, while the BIC provides moderate evidence supporting it. Since the BIC imposes a stronger penalty on additional parameters, the fact that both criteria favor the diffusive model suggests that the improvement obtained is statistically significant and not simply a consequence of increased model flexibility.

These results indicate that the inclusion of the diffusion contribution leads to an improvement in the phenomenological description of the combined BAO and Pantheon+ datasets. Moreover, the fact that the diffusive model remains favored even when spatial curvature is allowed to vary in the reference $\Lambda$CDM cosmology suggests that the effects introduced by diffusion cannot be trivially reproduced by curvature alone.

Finally, a critical consistency check for any model predicting a high local expansion rate ($H_0 \simeq 73.35$ km/s/Mpc) is the present age of the Universe, $t_0$. In standard $\Lambda$CDM cosmology, a higher $H_0$ typically reduces $t_0$ to roughly $12.8$ Gyr, creating a severe conflict with the age estimates of the oldest globular clusters ($t \gtrsim 13.5$ Gyr). In our framework, the age is determined by numerically integrating the modified expansion history, using the best-fit values of the parameters through the expression
\begin{equation}
    t_0 = \frac{1}{H_0} \int_{0}^{\infty} \frac{dz}{(1+z)E(z)}.
\end{equation}
Because the diffusion term grows significantly slower towards the past ($\beta \approx 0.503$) compared to standard matter, and due to the presence of negative spatial curvature ($\Omega_{k0} < 0$) it yields an estimated cosmic age of $t_0 \approx 13.61$ Gyr. This represents an interesting feature: it shows that unimodular diffusion can reconcile the locally observed high expansion rate with a Universe that remains old enough, thereby circumventing the cosmic age problem.

\subsection{Fits to the observational data}

\begin{figure}[h!]
\centering
\includegraphics[width=0.78\textwidth]{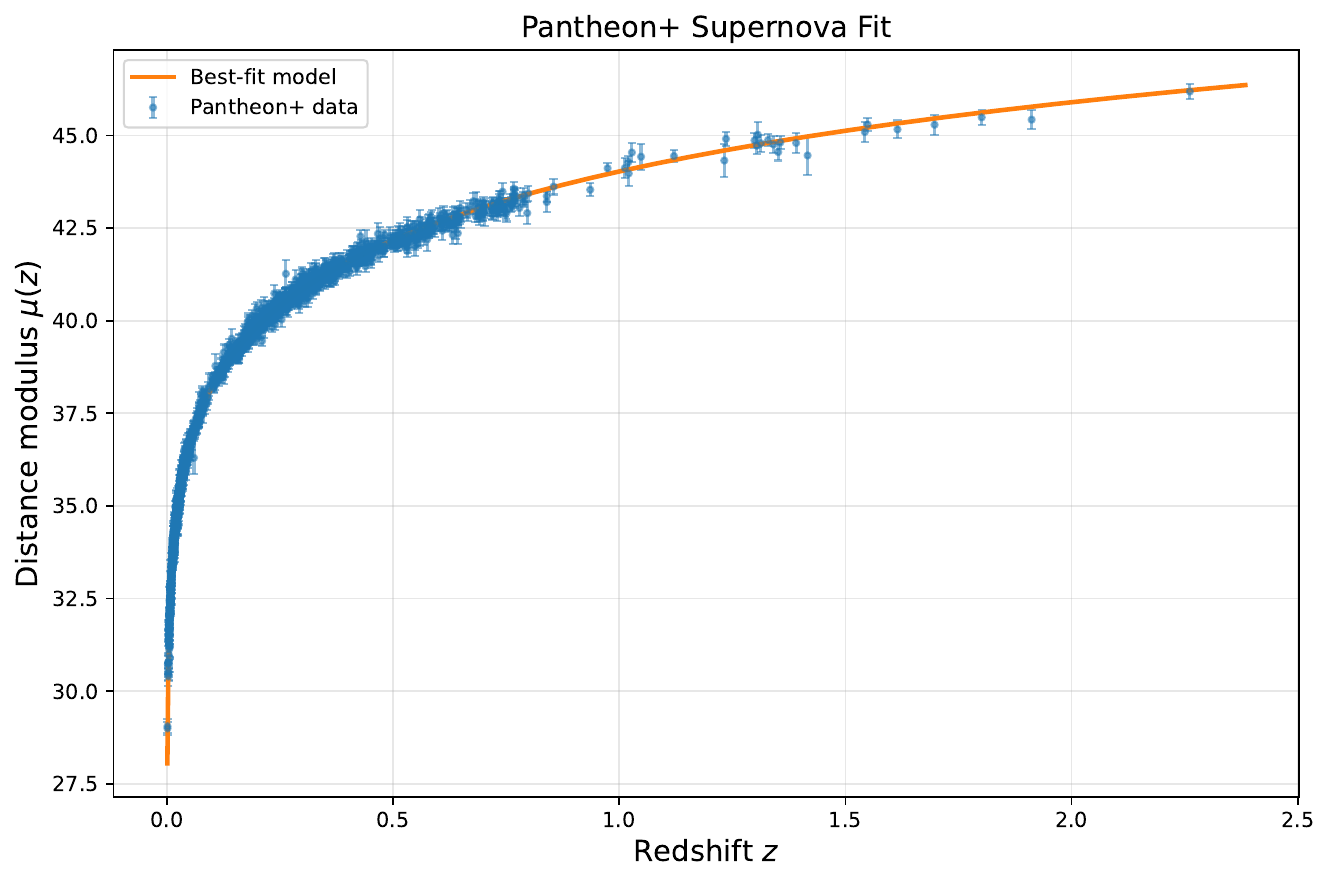}
\caption{Best-fit distance modulus for the diffusive dark matter model compared with the Pantheon+ Type Ia Supernova sample.}
\label{fig:pantheon_fit}
\end{figure}

\begin{figure}[h!]
\centering
\includegraphics[width=0.70\textwidth]{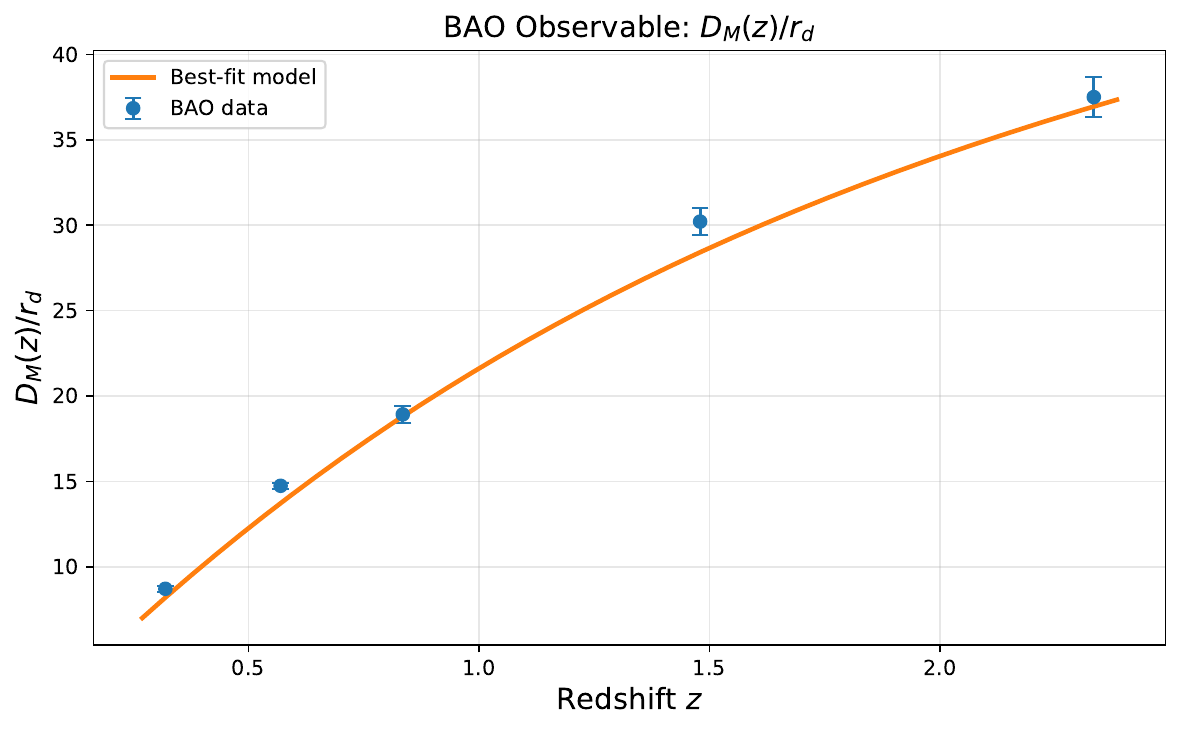}
\caption{Best-fit prediction for $D_M(z)/r_d$ compared with the BAO measurements used in the analysis.}
\label{fig:bao_dm}
\end{figure}

\begin{figure}[h!]
\centering
\includegraphics[width=0.70\textwidth]{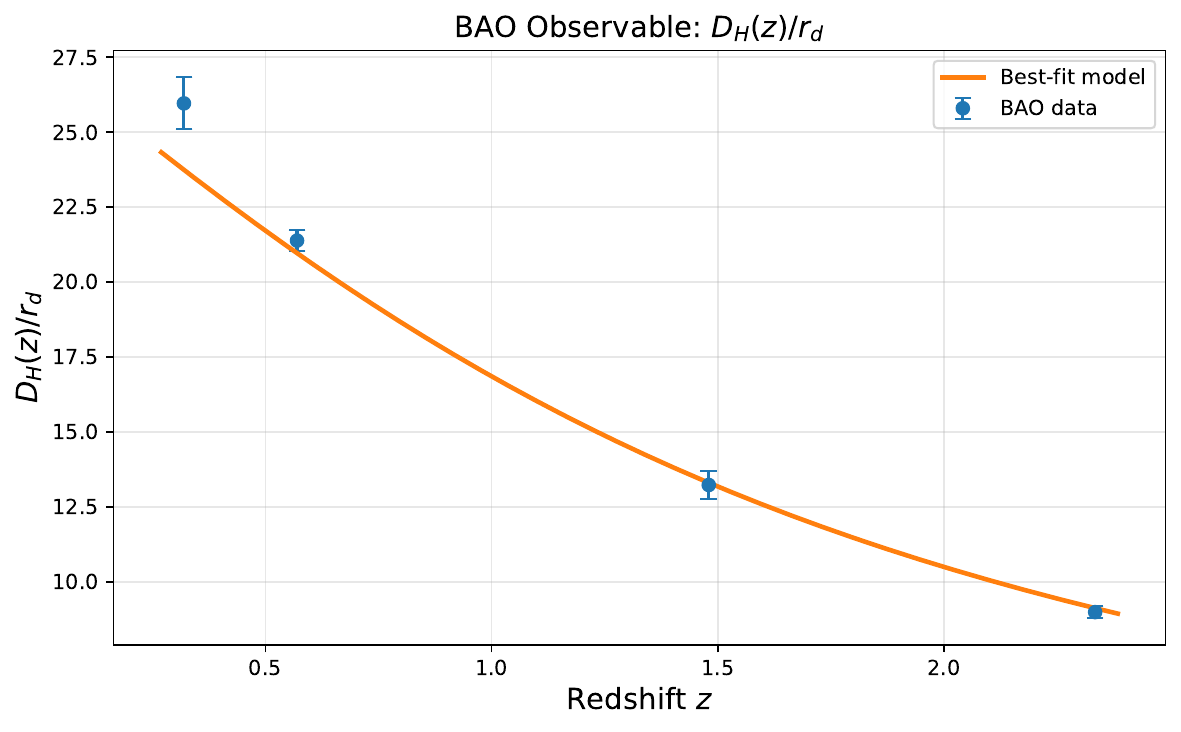}
\caption{Best-fit prediction for $D_H(z)/r_d$ compared with the BAO measurements used in the analysis.}
\label{fig:bao_dh}
\end{figure}

\begin{figure}[h!]
\centering
\includegraphics[width=0.70\textwidth]{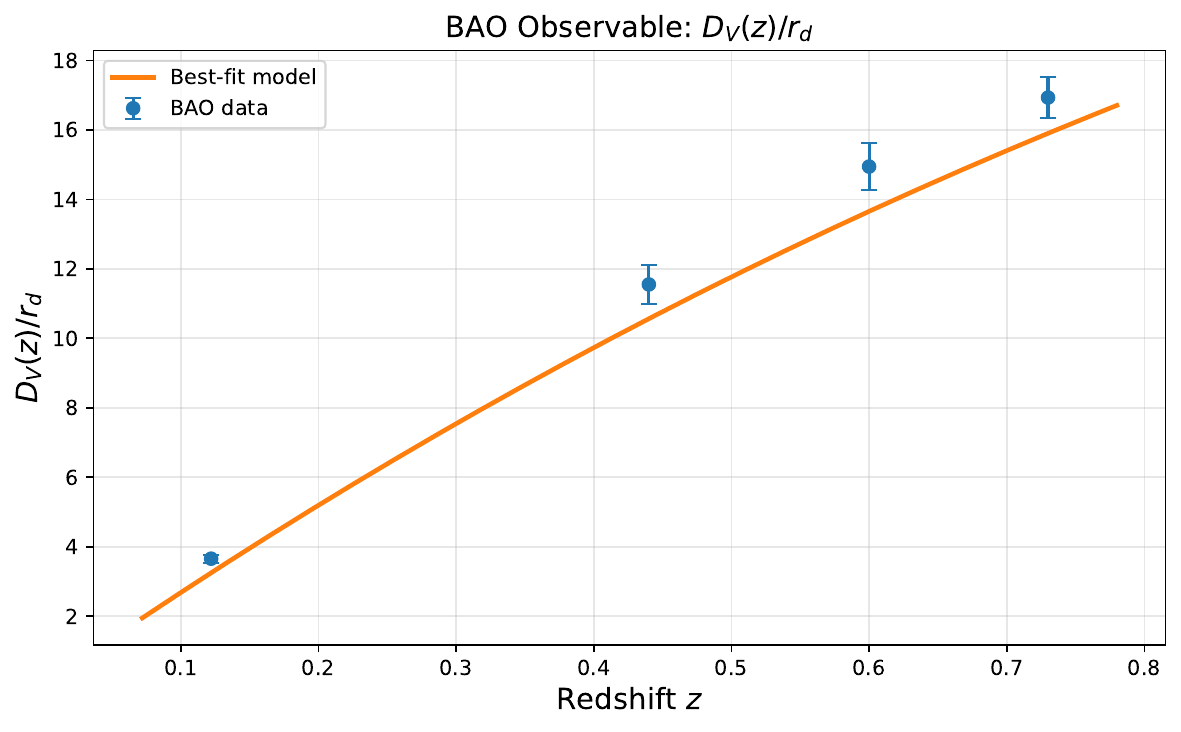}
\caption{Best-fit prediction for $D_V(z)/r_d$ compared with the BAO measurements used in the analysis.}
\label{fig:bao_dv}
\end{figure}

The fit to the Pantheon+ distance modulus is shown in Fig.~\ref{fig:pantheon_fit}. The theoretical prediction closely follows the supernova data over the full observed redshift range. The BAO constraints are shown separately for the three observables used in the likelihood. The corresponding best-fit curves are displayed in Figs.~\ref{fig:bao_dm}, \ref{fig:bao_dh}, and \ref{fig:bao_dv}.

The BAO constraints inherently depend on the dimensionless ratios $D_M(z)/r_d$, $D_H(z)/r_d$, and $D_V(z)/r_d$. Within the standard $\Lambda$CDM framework, imposing a fixed value of $r_d$ using early-universe CMB priors tightly constrains the subsequent expansion history, which typically worsens the discrepancy in $H_0$ when compared with local distance-ladder measurements. However, within our unimodular diffusion framework, the early-universe physics determining the absolute physical size of $r_d$ remains largely unaffected. The modifications to the expansion rate are governed by the diffusion scaling $(1+z)^\beta$ and the spatial curvature $(1+z)^2$. For our constrained value of $\beta \simeq 0.503$, the diffusion term grows significantly slower towards the past than both the matter sector $\propto (1+z)^3$ and the radiation sector $\propto (1+z)^4$. Consequently, both diffusion and curvature are highly subdominant at the drag epoch ($z_d \simeq 1059$). Because the pre-recombination dynamics are preserved, $r_d$ retains its validity as a standard ruler. The model provides a viable late-time mechanism capable of partially alleviating the Hubble tension through modified geometric projections at low redshift. The interplay between the slightly closed spatial geometry ($\Omega_{k0} \simeq -0.109$) and the diffusion dynamics provides sufficient flexibility in the comoving distances to simultaneously accommodate a higher local expansion rate ($H_0 \simeq 73.35$ km/s/Mpc) and the high-redshift BAO ratios. This demonstrates that unimodular diffusion successfully maps the early-universe sound horizon to the local Universe without requiring exotic pre-recombination physics.

\section{Concluding remarks}
\label{sec:vi}
We have investigated a cosmological scenario within the framework of Unimodular Gravity in which the non-conservation of the energy-momentum tensor gives rise to an effective energy diffusion function. In contrast with several previous diffusion proposals, whose thermodynamic behavior may become problematic in certain regions of the parameter space, we adopted a power-law Ansatz, and imposed from the outset the condition required by positive entropy production, namely $\beta Q_0>0$. This requirement restricts the physically viable sector of the model and ensures that the diffusion process is compatible with the second law of thermodynamics. In addition, we allowed for a non-vanishing spatial curvature contribution, which plays an important role in separating genuine diffusion effects from purely geometric distortions in the distance-redshift relation.

Focusing on a late-time dust-like regime, $\omega=0$, and setting the arbitrary unimodular integration constant to $\Lambda=0$, we derived the corresponding modified expansion law. The singular value $\beta=3$ was excluded from the allowed region, since at this point the diffusion and dust-like matter scalings become degenerate in the analytic solution for the energy density.

Using a joint analysis of Pantheon+ Type Ia Supernovae and BAO measurements, we obtained well-constrained posterior distributions for the parameters of the model. The diffusion exponent was constrained to $\beta=0.503^{+0.118}_{-0.126}$, while the present diffusion function $\Omega_{Q0}=0.725^{+0.037}_{-0.040}$ was found to be positive. This result is particularly relevant from the thermodynamic point of view, since the condition $\beta Q_0>0$ is naturally satisfied in the observationally preferred region. Moreover, the inferred value of $\beta$ implies that the effective diffusion contribution behaves as a quintessence-like component, with $\omega_{\rm eff}=\beta/3-1\simeq -0.832$, remaining outside the phantom regime. Therefore, the model avoids future Big Rip-type singularities associated with $\omega_{\rm eff}<-1$, while still providing a late-time accelerated behavior.

Another important outcome of the analysis is the role played by spatial curvature. We obtained $\Omega_{k0}=-0.109^{+0.076}_{-0.071}$, which points mildly towards a closed geometry, although still close to spatial flatness within the uncertainties. Thus, a moderately closed space emerges as the naturally preferred and adequate region to fit the observational datasets under a dynamic dark sector.
The fact that the preferred value of $\beta$ is clearly different from $\beta=2$ is also significant. If $\beta=2$, the diffusion term would scale as $(1+z)^2$ and would become observationally degenerate with the curvature contribution at the background level. Instead, the constrained value $\beta\simeq0.5$ indicates that the combined Pantheon+ and BAO data can distinguish the dynamical effect of unimodular diffusion from the purely geometric effect of spatial curvature.

The model also yields a high value of the Hubble constant, $H_0=73.350^{+0.221}_{-0.226}\ {\rm km\,s^{-1}\,Mpc^{-1}}$, which lies close to local distance-ladder determinations. At the same time, the modified expansion history gives a present cosmic age of approximately $t_0\simeq13.61\ {\rm Gyr}$. This provides an important consistency check, since high values of $H_0$ often lead to a younger Universe in standard cosmological models. In the present scenario, the slow growth of the diffusion term toward the past, together with the curvature contribution, allows the model to accommodate a high late-time expansion rate without generating a severe cosmic age problem.

We also compared the statistical performance of the diffusive model against a curved $\Lambda$CDM reference cosmology using information criteria. The differences $\Delta{\rm AIC}=-9.65$ and $\Delta{\rm BIC}=-4.21$, defined with respect to curved $\Lambda$CDM, indicate that the diffusive model is favored by both criteria. The AIC provides strong support for the diffusion scenario, while the BIC, despite imposing a stronger penalty on the additional parameter, still gives moderate evidence in its favor. This suggests that the improvements obtained by the model are not merely the result of additional parameter freedom, but rather reflect a genuine enhancement in the phenomenological description of the combined BAO and Pantheon+ datasets.

Additionally, in Appendix~\ref{sec:gpapp}, we present an independent consistency test based on a model-independent Gaussian Process reconstruction of $Q(z)$ using Cosmic Chronometer data. There, it can be seen that the reconstructed diffusion function remains positive over the observed redshift range, in agreement with the thermodynamic requirement of positive entropy production. Furthermore, the theoretical power-law Ansatz lies within the $2\sigma$ confidence region of the reconstruction over the full interval considered and matches the reconstructed behavior especially well near the present epoch. This agreement supports the viability of the power-law description.

Our results show that energy diffusion, when supplemented by spatial curvature and constrained by thermodynamic consistency, provides a physically viable and observationally competitive alternative to the standard $\Lambda$CDM framework. The model simultaneously provides a mechanism for late-time acceleration, a high value of $H_0$, a consistent cosmic age, and a statistically favored fit to background observations. Future explorations should include CMB distance priors, growth-rate data, and a more detailed perturbative analysis in order to determine whether the diffusion mechanism remains viable beyond the homogeneous background level. Such treatment remains necessary to determine the stability of the diffusion mechanism at the structure formation level.

\section*{Acknowledgments}
M.~Cruz and G. A. P. work has been supported by S.N.I.I. (SECIHTI-M\'exico). G. A. P. was supported by SECIHTI through the {\it Estancias Posdoctorales por México 2023(1)} program. S.~Lepe acknowledges the FONDECYT grant N°1250969, Chile.

\appendix
\section{Gaussian Process Reconstruction}
\label{sec:gpapp}
To perform a model-independent assessment of the energy diffusion process, we employ Gaussian Processes (GPs) \cite{gp}. A GP describes a distribution over functions and is thus a generalization of Gaussian distributions to function space. The analysis is fully Bayesian; we start with a prior for the function distribution and combine it with the likelihood of observing the data, which leads to a posterior function distribution. A Gaussian process $f(z)$ is completely specified by its mean function $\mu(z)$ and covariance function $k(z, \tilde{z})$. In accordance with the formalism described in (citar), we define a distribution over functions by assuming a zero a priori mean function, $\mu(z)=0$, and adopting a Mat\'ern $\nu=5/2$ covariance function $k(z, \tilde{z})$ to specify the relationship between function values at different redshifts. The covariance function depends on the hyperparameters, such as the characteristic length scale $l$ and the signal variance $\sigma_f$. These hyperparameters do not specify the exact form of the function, but rather characterize its {\it bumpiness}, and they are trained by maximizing the marginal likelihood. For the Gaussian Process reconstruction, we utilize a comprehensive set of 32 Cosmic Chronometer (CC) measurements, which provide direct, model-independent estimates of the Hubble rate $H(z)$ by measuring the age difference between passively evolving galaxies \cite{cc}. These data points are crucial as they do not rely on an assumed distance-redshift relation. The observational noise is incorporated into the GP framework through the covariance matrix of the data, $C = \text{diag}(\sigma_i^2)$, assuming that the errors follow a Gaussian distribution. Furthermore, to provide a rigorous comparative analysis, the free parameters of our theoretical diffusion model ($\Omega_{Q0}$ and $\beta$) were obtained from a separate joint likelihood analysis. This analysis combines the Pantheon+ Type Ia Supernova sample, providing luminosity distance measurements, and BAO data, which serve as a standard ruler at different cosmic epochs. A key advantage of this method is that the derivative of a Gaussian process is again a Gaussian process. This allows us to reconstruct the Hubble parameter $H(z)$ and its first derivative $H'(z)$ directly from the Cosmic Chronometer data. By evaluating the joint predictive distribution and its associated covariances, we propagate the uncertainties using a Monte Carlo sampling approach to robustly determine the non-parametric diffusion function $Q(z)$ and its corresponding confidence levels. In the context of UG, the energy diffusion function $Q(z)$ is coupled to the modified expansion history. By using the field equations for a FLRW universe with spatial curvature $k$, we can express the diffusion process in a purely kinematic form. Given the dimensionless Hubble rate $E(z) = H(z)/H_0$ and its derivative $E'(z)$, both obtained through the GP reconstruction, the dimensionless diffusion function is determined by:
\begin{equation}
\frac{Q(z)}{H_0^2} = 3E^2(z) - 2(1+z)E(z)E'(z) - \Omega_{k0}(1+z)^2,
\label{eq:Q_reconstructed}
\end{equation}
where $\Omega_{k0}$ represents the dimensionless spatial curvature density parameter. This kinematic strategy, which extracts interaction or diffusion contributions directly from the modified continuity equation together with derivatives of the expansion rate, has been demonstrated to be a highly robust, non-parametric method \cite{cai}. This formulation ensures that the reconstructed $Q(z)$ is independent of any dark energy parametrization, providing a direct observational test for the thermodynamic consistency of the unimodular framework.

\subsection{Results and Discussion: Observational Validation of the Diffusion Ansatz}
\label{sec:discussion}

\begin{figure}[h!]
\centering
\includegraphics[width=1\textwidth]{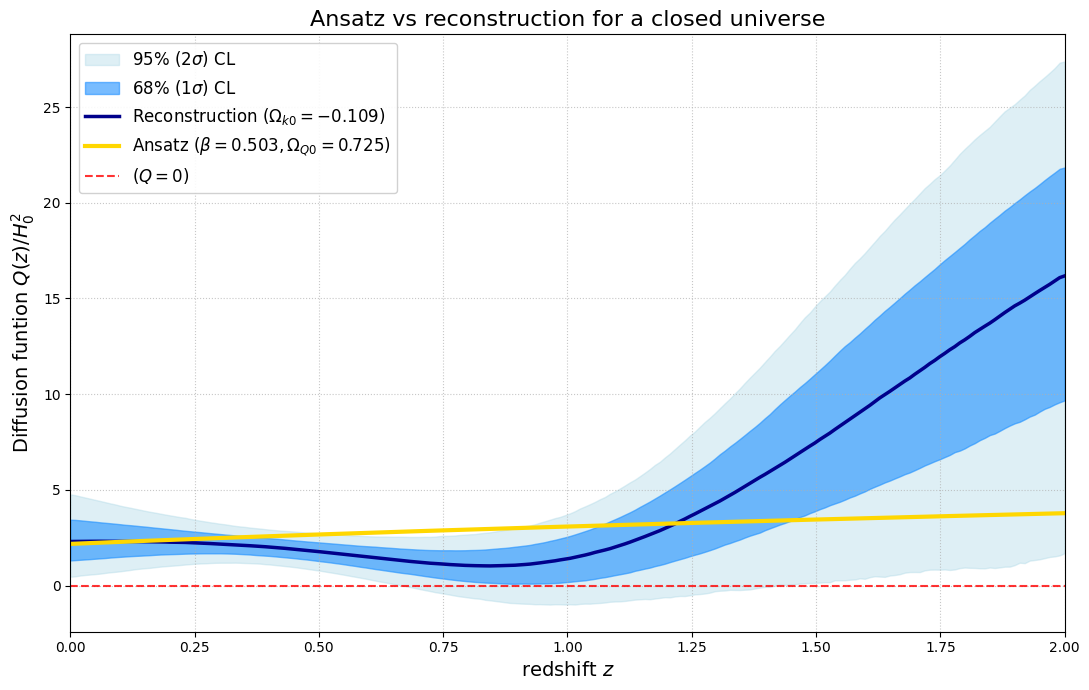}
\caption{Comparison of the power-law Ansatz for $Q(z)$ with the GP-reconstructed function for a closed universe, using our best-fit curvature value ($\Omega_{k0} = -0.109$).}
\label{fig:reconstruction}
\end{figure}

The model-independent reconstruction of the diffusion function $Q(z)$ serves as a key reference point for assessing the validity of the theoretical power-law Ansatz introduced in this study. In Figure \ref{fig:reconstruction}, we present the dimensionless diffusion function $Q(z)/H_0^2$ reconstructed via GPs for a closed universe using our best-fit curvature parameter ($\Omega_{k0} = -0.109$). The reconstructed mean function is shown alongside its 68\% ($1\sigma$) and 95\% ($2\sigma$) confidence level (CL) regions. Crucially, the GP reconstruction reveals that the diffusion remains strictly positive ($Q(z) > 0$) across the entire observed redshift range $z \in [0, 2]$. This result acts as an independent observational confirmation of the thermodynamic consistency of our model, naturally satisfying the requirement of positive entropy production ($\beta Q_0 > 0$) without prior theoretical imposition. We consider our theoretical Ansatz, $Q(z)/H_0^2 = 3 \Omega_{Q0} (1+z)^\beta$, using the constrained parameters $\beta = 0.503$ and $\Omega_{Q0} = 0.725$. As observed in the figure, the theoretical Ansatz lies entirely within the 95\% CL ($2\sigma$) region of the GP reconstruction for the whole redshift range. There is a slight deviation from the 68\% CL ($1\sigma$) region at intermediate redshifts ($0.25 \lesssim z \lesssim 1.4$). This moderate tension is an expected consequence of combining different datasets; our Ansatz is calibrated primarily on the local expansion rate favored by the Pantheon+ sample and BAO, whereas the GP is trained on Cosmic Chronometer measurements, which typically favor lower $H_0$ values. Despite these methodological differences, the Ansatz remains statistically compatible with the model-independent reconstruction at the $2\sigma$ level globally.
It is noteworthy that the GP reconstruction and the theoretical power-law Ansatz coincide almost perfectly in the low-redshift limit, specifically at $z \simeq 0$. As discussed in \cite{gp}, Gaussian Processes are highly sensitive to the local data density, often resulting in a {\it sweet-spot} where reconstruction errors are minimized. The fact that our theoretical model, calibrated on Pantheon+ and BAO datasets, aligns with the model-independent CC reconstruction at $z \simeq 0$ suggests that the present-day diffusion amplitude $\Omega_{Q0}$ is a robust physical quantity. This agreement at the anchor point of the expansion history provides additional support for the thermodynamic consistency of the Unimodular framework and validates the power-law scaling as a reliable description of late-time cosmic evolution.

\end{document}